\documentclass[12pt,preprint]{aastex}
\shortauthors{Carciofi et al. }
\shorttitle{}

\begin{document}

\title{On the Determination of the Rotational Oblateness of Achernar}

\author{
A.~C.~Carciofi \altaffilmark{1},
A. Domiciano de Souza\altaffilmark{2},
A. M. Magalh\~aes\altaffilmark{1},
J. E. Bjorkman\altaffilmark{3},
F.~Vakili\altaffilmark{2}
}

\altaffiltext{1}{Instituto de Astronomia, Geof{\'\i}sica e Ci{\^e}ncias
Atmosf{\'e}ricas, Universidade de S\~ao Paulo, Rua do Mat\~ao 1226,
Cidade Universit\'aria, S\~ao Paulo, SP 05508-900, Brazil}
\altaffiltext{2}{LUAN, Universit{\'e} de Nice-Sophia Antipolis (UNSA),
CNRS, Observatoire de la C{\^o}te d'Azur (OCA), 28 avenue de Valrose,
Parc Valrose, F-06108 Nice - France} \altaffiltext{3}{University
of Toledo,  Department of Physics \& Astronomy, MS111 2801 W.
Bancroft Street Toledo, OH 43606 USA}
\begin{abstract}

The recent interferometric study of Achernar, leading to the
conclusion that its geometrical oblateness cannot be explained in
the Roche approximation, has stirred substantial interest in the community, in view
of its potential impact in many fields of stellar astrophysics.
It is the purpose of this paper to reinterpret the interferometric
observations with a fast rotating, gravity
darkened central star surrounded by a small equatorial disk, whose
presence is consistent with contemporaneous spectroscopic data.
We find that we can only fit the available data assuming a critically rotating central star.
We identified two different disk models that 
simultaneously fit the spectroscopic, polarimetric, and
interferometric observational constraints: a tenuous disk in
hydrostatic equilibrium (i.e., with small scaleheight) and a smaller, scaleheight enhanced disk. 
We believe that these relatively small disks correspond to the
transition region between the photosphere and the circumstellar
environment, and that they are probably perturbed by some photospheric mechanism.
The study of this interface between photosphere and circumstellar
disk for near-critical rotators is crucial to our understanding of
the Be phenomenon, and the mass and angular momentum loss of stars
in general. This work shows that it is nowadays possible to
directly study this transition region from simultaneous
multi-technique observations.

\end{abstract}


\section{Introduction \label{introduction}}

Interferometry greatly increased our knowledge of the Be Star
Achernar ($\alpha$ Eri, HD10144). \citet[][hereafter D03]{dom03}
observed Achernar during fall 2002 with the ESO-VLTI/VINCI instrument \citep{ker03} 
in the $K$ band and found that the star is highly
oblate. By converting the individual visibilities into equivalent
uniform disk angular diameters they derived a ratio between the
maximum and minor elongation axis $a/b=1.56\pm0.05$. Later,
\citet{ker06} detected a tenuous polar wind and very recently
\citet{ker07} found that Achernar has a main-sequence, lower mass
companion.

The available interferometric data, together with the abundance of
data from the literature, forms a body of information that is
probably unrivaled by that of any other Be star. There is,
however, much that is still unknown about the star and its
circumstellar environment. As pointed out by D03, the
determination of the actual stellar oblateness from the
interferometric data is not a simple task, since there are so many
unknowns involved. In their original work D03, based on
simultaneous spectroscopic data, assumed that there was no
circumstellar material at the time of the observations, and that
the observed shape was purely photospheric in origin. Their
modeling of the observations with a rotationally deformed, gravity darkened star, 
assuming the Roche approximation of uniform rotation and centrally condensed
mass,
led to the conclusion that the stellar flattening required to explain the
observations exceeds that of a critical rigid rotator .

Recent theoretical developments in the theory of differentially
rotating stars (e.g., \citealt{jac04}) show that they can have
equatorial radii larger than the Roche limit, which is physically
compatible with the interferometric results. However,
\citet{vin06} demonstrated the existence of a tiny, yet non
negligible emission in H$\alpha$ at the time of the VLTI observations,
an indication that there was some circumstellar material around
Achernar, probably associated with a small rotating disk.
Furthermore, \citet{vin06} showed, from temporal analysis of
profile variations of the line He I $\lambda6678$ \AA, that orbiting gas clouds
are a very frequent feature of Achernar, even during its quiescent
phase.

The presence of this
circumstellar material raises the question of how much it can
contribute to the observed visibilities. In this paper we
investigate how the presence of a small disk around Achernar may
alter to the interferometric signal, and the effects this may
have on determination of fundamental parameters such as stellar
rotation rate and stellar flattening.

\section{Models  \label{models}}

To compute the emergent spectrum of the system we use the computer
code HDUST \citep{car06a,car06b}. This code solves the coupled
problem of the non-local thermodynamic equilibrium (NLTE) and
radiative equilibrium for arbitrary three-dimensional envelope geometries,
using the Monte Carlo method. Given a prescription for the star
(rotation rate, shape, flux distribution, effective temperature,
luminosity, etc.) and for the circumstellar gas (density and
velocity distribution) the code calculates the NLTE Hydrogen level
populations and the electron kinetic temperature. Once those
quantities are determined, the emergent spectrum (spectral energy
distribution, line profiles, polarization, synthetic images, etc.)
is calculated.

For the present work, we modified HDUST to treat non-spherical,
gravity darkened stars. We assume the stellar shape to be an
spheroid, which is a reasonable approximation for the shape of a
rigidly rotating star  \citep{fre05}. Given the stellar rotation
rate, $\Omega/\Omega_\mathrm{crit}$, the ratio between the
equatorial and polar radii is determined from the Roche
approximation for the stellar surface equipotentials
\citep{fre05}. For the gravity darkening we use the standard von
Zeipel flux distribution \citep{von24}, according to which
$F(\theta) \propto g_\mathrm{eff}(\theta) \propto
T_\mathrm{eff}^4(\theta)$, where $g_\mathrm{eff}$ and
$T_\mathrm{eff}$ are the effective gravity and temperature at
stellar latitude $\theta$. In HDUST, the star is divided into a
number of latitude bins (typically 50) which have an associated
$T_\mathrm{eff}$ and $g_\mathrm{eff}$ and emit with an spectral
shape given by the corresponding Kurucz model atmosphere
\citep{kur94}.

For the disk density distribution we assume the following expression
\begin{equation}
\rho(\varpi,z)=\frac{\rho_0R_\mathrm{eq}}{\sqrt{2\pi}H}
\left(\frac{R_\mathrm{eq}}{\varpi}\right)^{2} \exp\left(-z^2/H^2\right),
\end{equation}
which is similar to the density distribution of an isothermal
viscous decretion disk in vertical hydrostatic equilibrium (e.g.
\citealt{car06b}). In the above equation, $\varpi$ is the radial
distance in cylindrical coordinates, $R_\mathrm{eq}$ is the
equatorial radius of the star and $\rho_0$ is the disk density
scale. We write the disk vertical scaleheight, $H$, as
\begin{equation}
H=H_0 \left(\varpi/R_\mathrm{eq}\right)^{1.5},
\label{eq:H}
\end{equation}
where $H_0$ is the scaleheight at the base of the disk. For
isothermal disks in vertical hydrostatic equilibrium $H_0=aV_{\rm
crit}^{-1}R_\mathrm{eq}$, where  $a$ is the sound speed and the
critical velocity, $V_{\rm crit} \equiv \left(GM/R_{\rm eq}\right)^{1/2}$, is the Keplerian orbital speed at 
the stellar surface.

We chose a set of fixed stellar parameters, listed in
Table~\ref{tab:star}. The value for the polar radius came directly
from the interferometry, once we fixed the stellar inclination
angle to be $65\deg$ after the recent work of \citet{car07}. It is
worthy noting, however, that the results we show below are little
affected if we allow the inclination angle to vary within a
reasonable range (say, $\pm 5\deg$). 
The value for the stellar luminosity was obtained by fitting the available photometric data (see caption of Figure~\ref{fig:sed}).

Each model of the star plus
disk system has 4 free parameters: (1)  the stellar angular
rotation rate, $\Omega/\Omega_\mathrm{crit}$, (2) the disk density
scale, $\rho_0$, (3) the disk outer radius, $R_\mathrm{d}$, and
(4) the disk scaleheight, $H_0$.

We have two well-known observational constraints, viz., the 2002
interferometric observations (D03) and the 2002 H$\alpha$ line
profile (\citealt{vin06}, fig. 12).
Unfortunately, no
contemporaneous measurement of the linear polarization exists, but
we can impose, with a reasonable level of confidence, an upper
limit for the polarization based on the recent results of
\citet{car07}. In that paper, the results of a
polarization monitoring of Achernar, carried out between 2006 July and
 November, is reported. Those measurements were taken during a period in
which Achernar was active, with a tenuous circumstellar disk, as
indicated by the weak H$\alpha$ emission (\citealt{car07}, fig. 3). We adopt as an upper limit for the
polarization in our modeling the lowest value reported in
\citet{car07}, which was 0.12\% in the $B$ band. As we shall see
below, fixing this upper limit for the polarization level has
important consequences for the modeling. 

\section{Results  \label{results}}

Our modeling procedure is as follows. For a given set of
$\Omega/\Omega_\mathrm{crit}$, $\rho_0$, $R_\mathrm{d}$ and $H_0$
we calculate synthetic images in the $K$ band, centered at
$\lambda=2.15\;\mu\rm m$, the H$\alpha$ line profile, the
continuum polarization and the spectral energy distribution (SED).
The model amplitude visibilities where obtained from the Fourier
transform moduli of the model images at $2.15\;\mu\rm m$,
normalized by 
the total $K$ band flux of the synthetic image.

Since the H$\alpha$ emission from October 2002 is very small and
the true photospheric profile of Achernar is not known, we model,
instead of the observed profile, the residual emission profile of
\citet{vin06}, fig. 12. This profile was obtained by subtracting
from the October 2002 observations the average profile of the 1999
period, which is believed to be purely photospheric. The
equivalent width (hereafter EW) of this emission profile is
$-0.29$ \AA.

In Figure~\ref{fig:models} we show the visibility curves along the
polar and equatorial directions together with the corresponding
$K$ band images for five representative models that were chosen to 
illustrate different aspects of our solution. 
In Table~\ref{tab:models} we list the model parameters, along with 
some model results, such as the $B$ band polarization level, the
H$\alpha$ EW and the $K$ band flux excess, $E_K$, which is defined
as $F^\lambda/F^\lambda_\star-1$, where $F^\lambda_\star$ is the
stellar flux without the disk at wavelength $\lambda$.

Let us begin discussing models 1 to 3, for which the star was
assumed to be rotating critically. We have used
$\Omega/\Omega_\mathrm{crit} = 0.999999$
($V_\mathrm{eq}/V_\mathrm{crit}=0.9993$) instead of 1 to avoid the
unphysical situation of having $T_{\rm eff} = 0$ in the stellar
equator. Model 1 corresponds to a small and relatively dense disk,
with $H_0$ given by Eq.~\ref{eq:H}, i.e., model 1 corresponds to a
disk in vertical hydrostatic equilibrium. For this model the
visibility curves matches very well the observations and the
polarization is within the adopted limit, but the H$\alpha$
emission is very weak, as a result of the small disk size.

One way to increase the H$\alpha$ emission is to make the disk larger.
For model 2 we adopted a larger value for  $R_d$, but the density had to be decreased in order to keep the polarization within the adopted limit.
This model reproduces well both the visibility curves and the H$\alpha$ EW. 

Another way of increasing the line emission is to raise the disk scaleheight.
If one raises $H$ by a factor of, say, $k$, the disk density scale must be decreased by a factor $k^{1/2}$ to keep the polarization approximately constant.
Model 3 corresponds to our best scaleheight enhanced model for a critically rotating star.
The visibility curves and the  H$\alpha$ EW are well reproduced by the model, and the polarization is within the adopted upper limit.

We also studied models with lower values of $\Omega/\Omega_\mathrm{crit}$. In this case, our attempts to fit simultaneously the visibility curves, polarization and H$\alpha$ EW were unsuccessful. 

Models 4 and 5 are examples of the results we have obtained for
sub-critical stars. Model 4 corresponds to a small and dense disk
in  vertical hydrostatic equilibrium. This model reproduces quite
well the visibilities, but have too large $B$ band polarization and no H$\alpha$ emission. A
scaleheight enhanced model with lower density, such as model 5,
have better values for polarization and H$\alpha$ EW, but does not
have enough IR flux in the equator to account for the observed
size.

Our lack of success on fitting the observations with sub-critical
models is a direct result of two physical effects, both related to
the presence of gravity darkening. For the critical models 1 to 3,
the bolometric stellar flux at the equator is about 1100 times
lower than the flux at the pole. Since the disks are relatively
small and the star oblate, light from the pole cannot reach the
disk and, as a result, the net scattered flux is small. As
$\Omega/\Omega_\mathrm{crit}$ decreases, the flux at the equator
rises, and the star becomes less oblate. Both mechanisms cause a
significant increase in the scattered flux and, thus, in the
polarization level.

\section{Discussion  \label{conclusions}}

\begin{figure}[bp]
\plotone{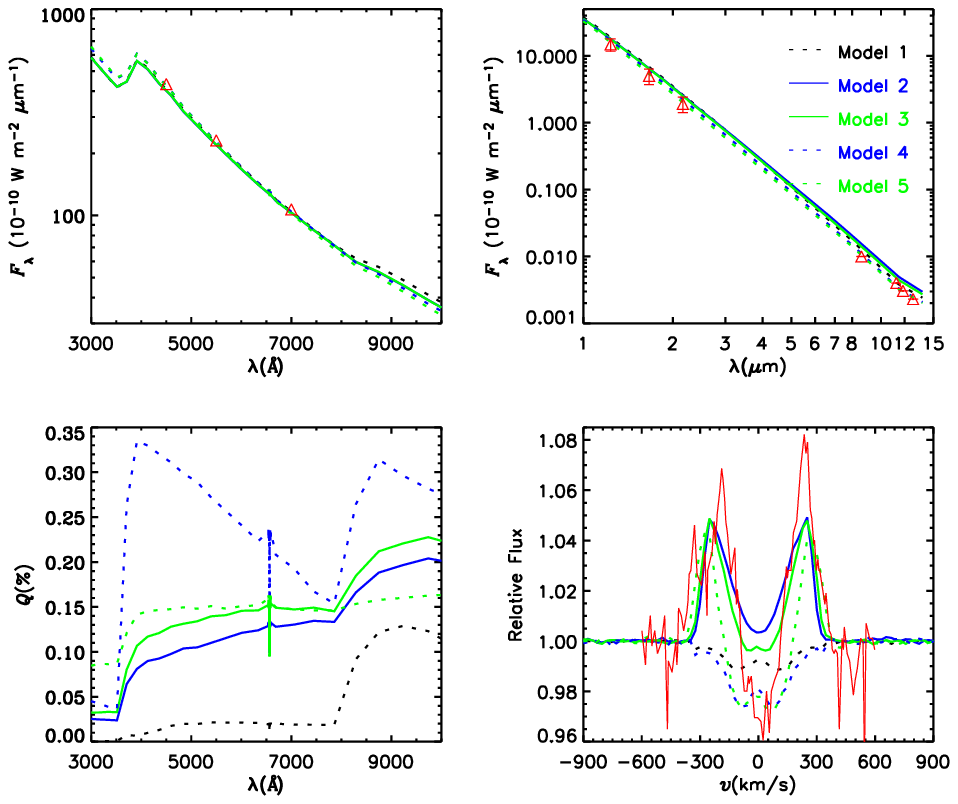} \figcaption[]{
Emergent spectrum for models 1 to 5. 
Top left: visible SED. The red triangles correspond to BVR photometry from the NOMAD catalog \citep{nomad}, for which no observational errors were available.
Top right: IR SED. The JHK photometry is from the 2MASS catalog \citep{2mass} and the mid-IR photometry is from \citet{ker07}. The observational errors for the mid-IR data is of the order of 2\%. 
Bottom left: continuum polarization.
Bottom right: H$\alpha$ emission profile. The red curve corresponds to the residual emission profile of \citet{vin06}.
\label{fig:sed}}
\end{figure}

\begin{figure}[bp]
\plottwo{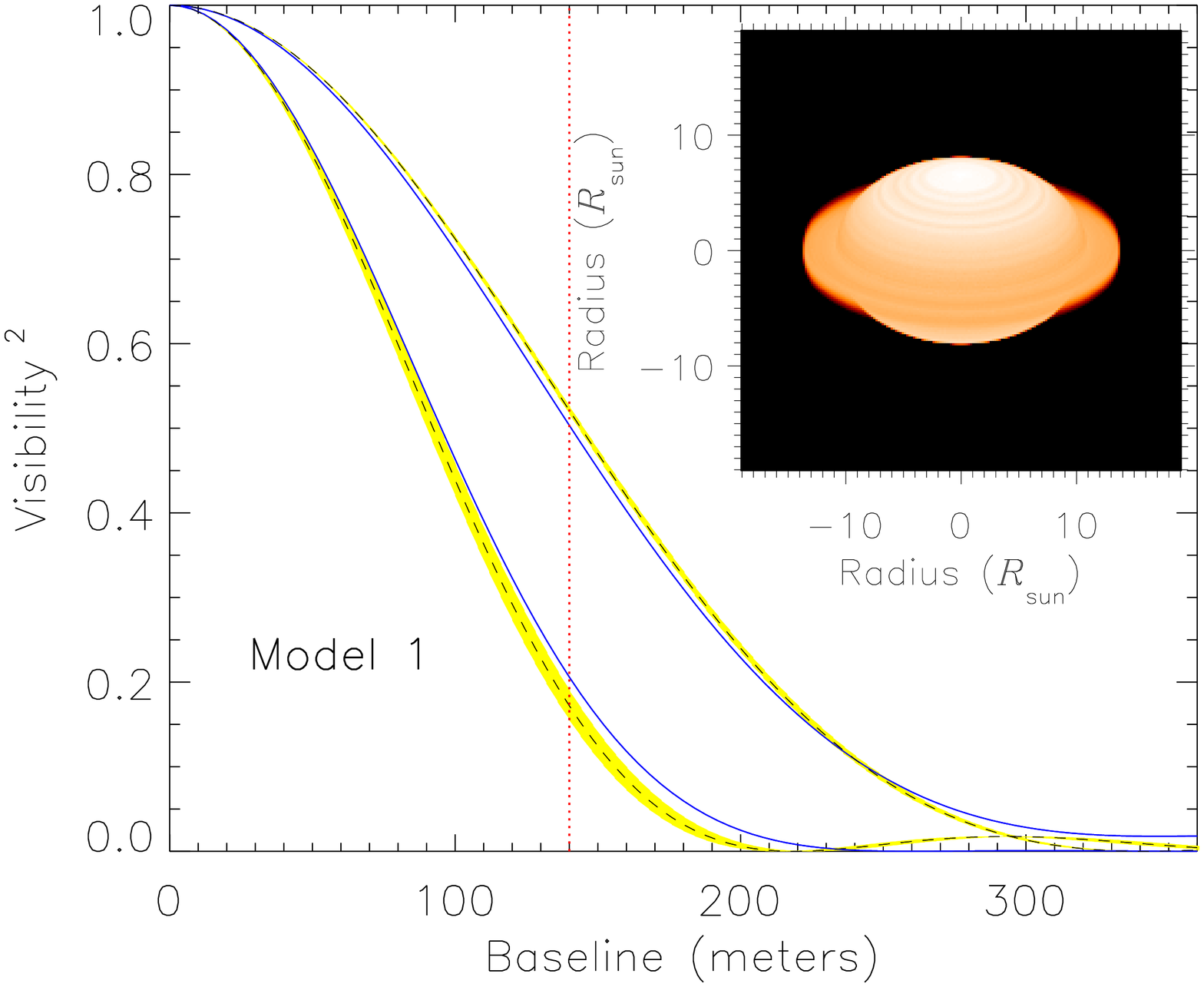}{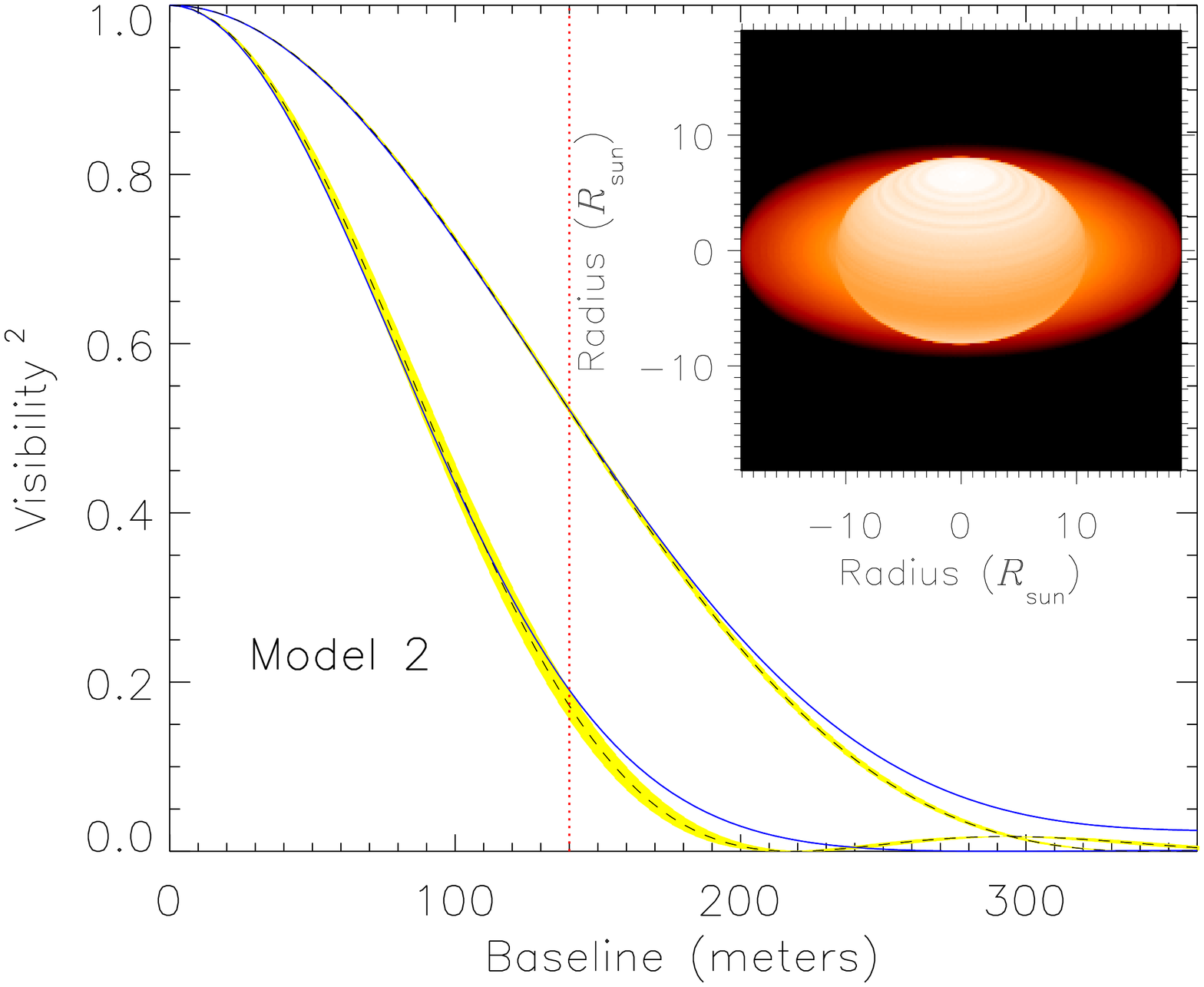}
\plottwo{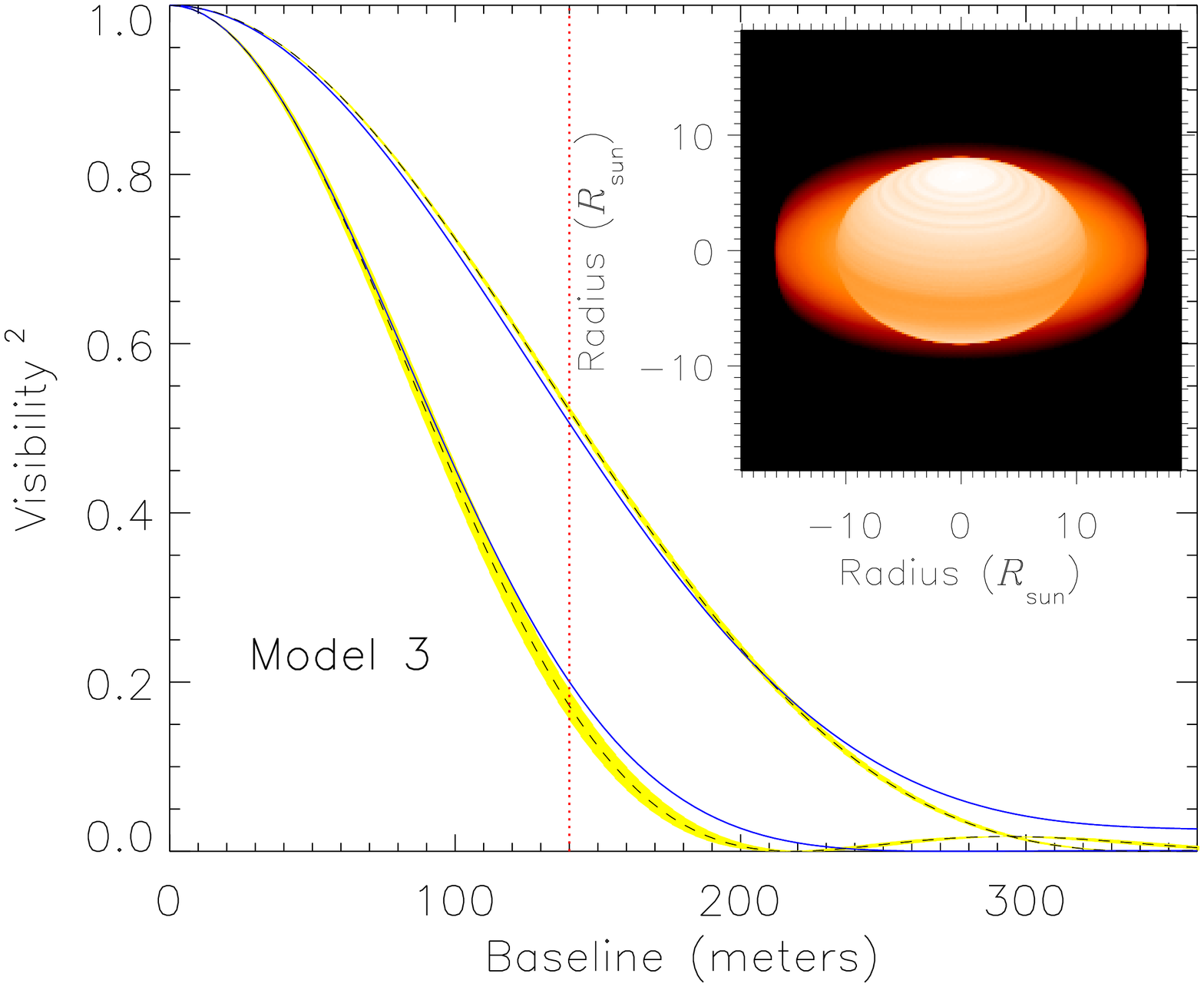}{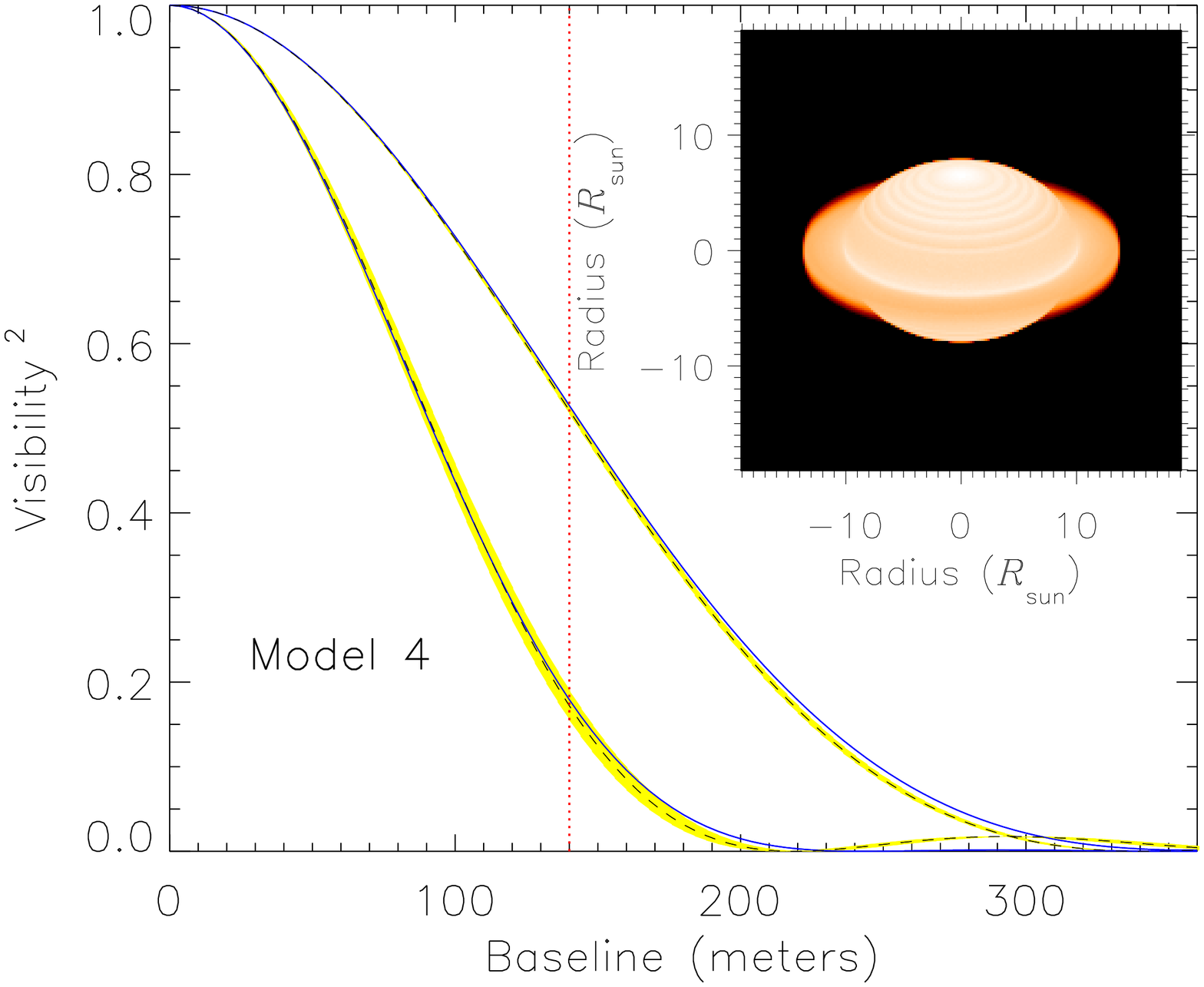}
\plottwo{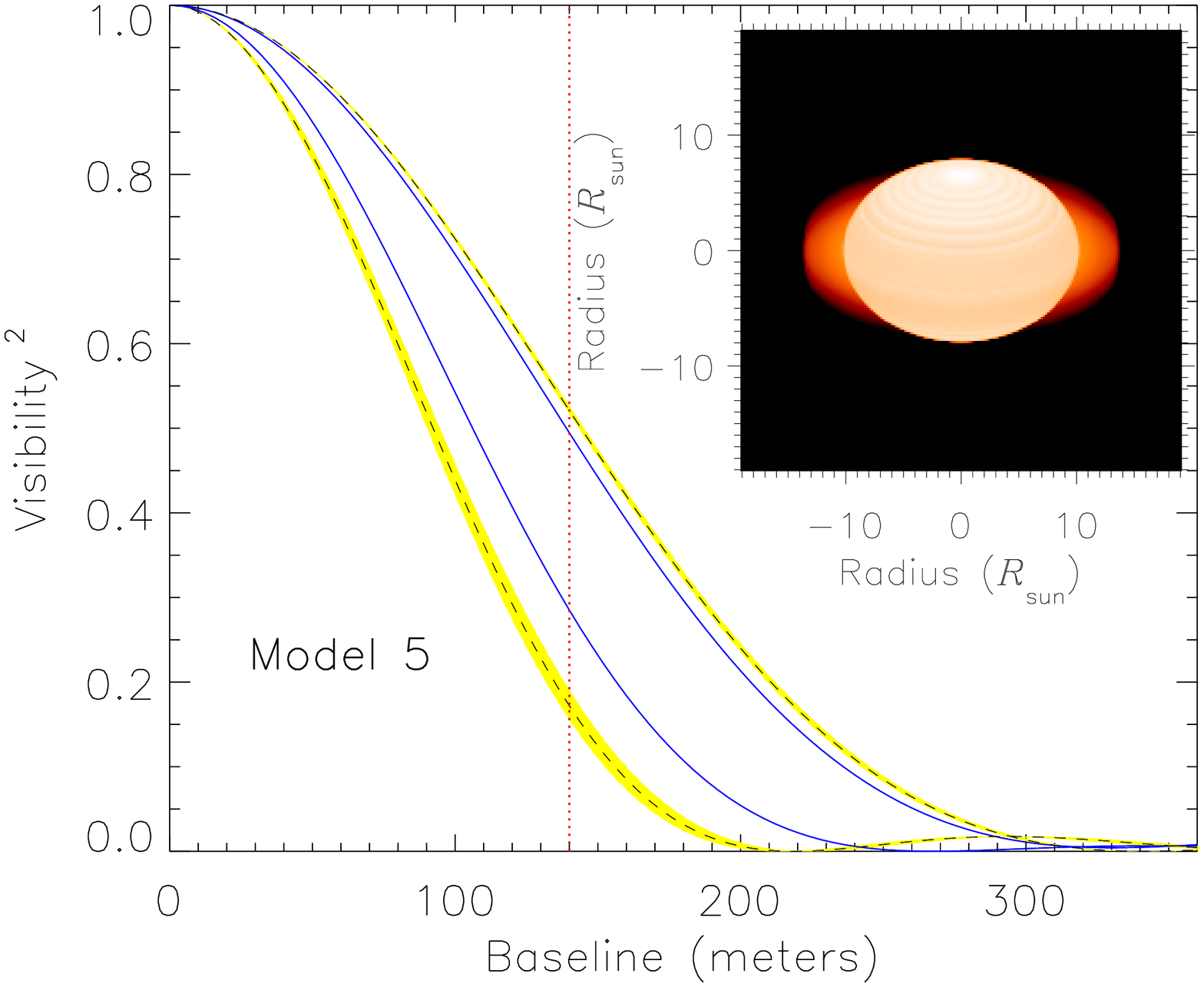}{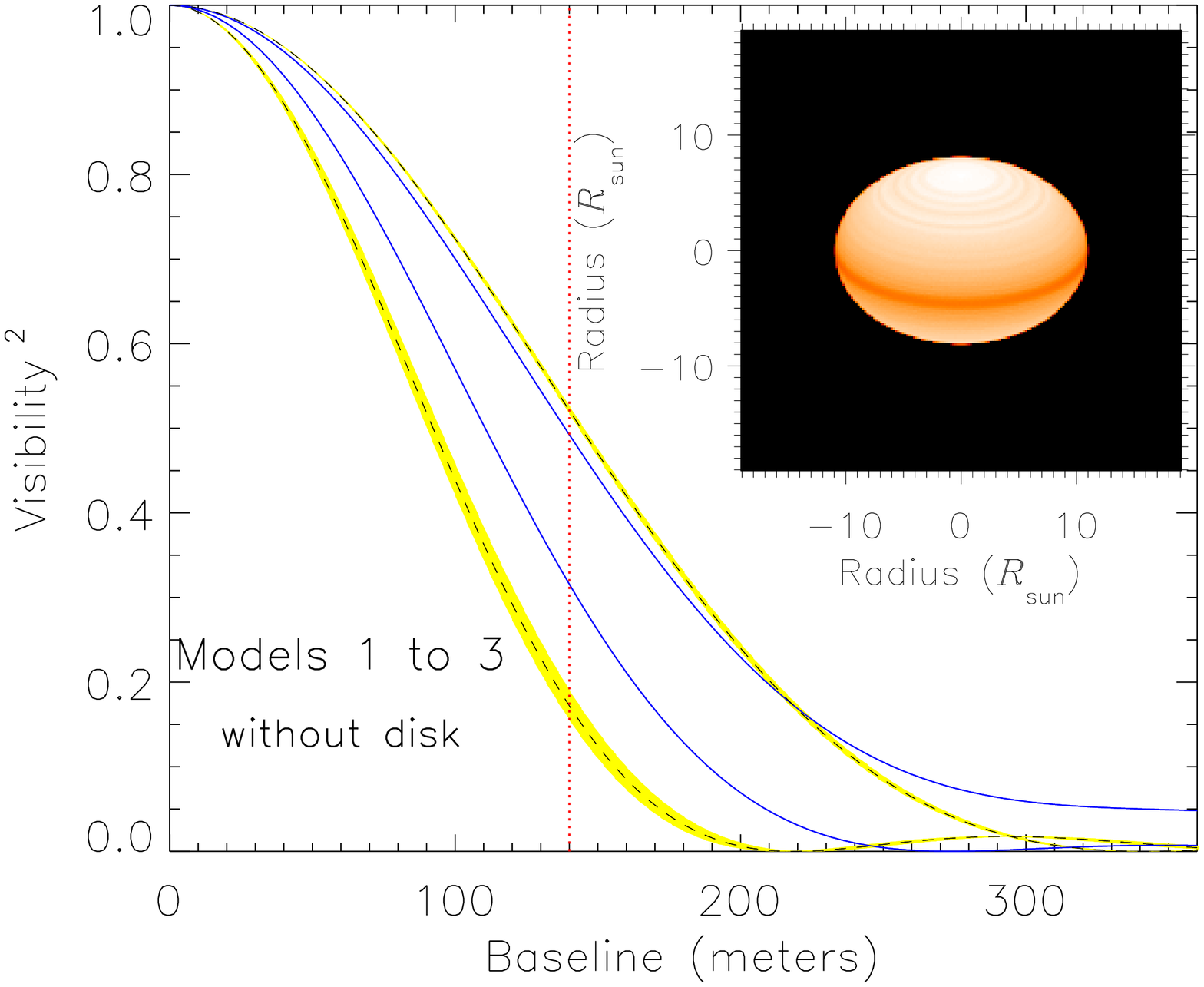}
 \figcaption[]{Squared visibilities for the models of Table~\ref{tab:models} (solid lines), along the polar (upper curves) and equatorial (lower curves) directions. The dashed lines represent the $K$ band visibilities for uniform disk (UD) angular diameters of 1.62~mas and 2.53~mas derived by D03; the corresponding $\pm1$-sigma uncertainties are shown as light-color bands. These UD diameters indicate the maximum (equatorial direction) and minimum (polar direction) sizes measured on Achernar in the $K$ band with VLTI/VINCI. The vertical dotted lines indicate the maximum baseline available from the VLTI data. The insets show the model image in the $K$ band, in logarithmic scale. The bottom right plot shows the squared visibilities for a gravity darkened Roche star at near-critical velocity (models 1 to 3) without the disk, and indicates, as shown by D03, that such model does not reproduce the observations. \label{fig:models}}
\end{figure}

The presence of a small disk around Achernar is a realistic
possibility to explain the interferometric observations. As shown
by D03 (see also Figure~\ref{fig:sed}), Roche models without a disk cannot
account for the observed aspect ratio.

The combined constraints imposed by the interferometry,
spectroscopy and the adopted upper limit for the polarization have
allowed us to impose very narrow limits on the model parameters,
within our assumption of a rigidly rotating star in the Roche
approximation surrounded by a small disk. The main result is that the star must be rotating
very close to critical, since all of our models with
$\Omega/\Omega_\mathrm{crit} < 0.992$ that successfully reproduced
both the visibilities and the $H\alpha$ EW (e.g. model 5)  had too
large polarization levels, for the reasons discussed above.

We have found two critical models that reproduce equally well the observations: a large and dense disk in hydrostatic equilibrium (i.e., geometrically thin; model 2) and a small and more tenuous disk with enhanced scaleheight (i.e., geometrically thick; model 3).
Because the lack of contemporaneous polarization measurements, the model parameters shown in Table~\ref{tab:models} have a degree of uncertainty that is
difficult to estimate. This stresses the importance of having simultaneous multi-technique data.

The current paradigm for Be star disks is that of a geometrically thin viscous
Keplerian disk in vertical hydrostatic equilibrium. This paradigm
has been corroborated by several recent studies
\citep[e.g.][]{car06b,mei07}.

When a star is close to critical rotation, the stellar photospheric material near the                  
equator, which is weakly bound to the star because of the strong
centrifugal force, can eventually escape the star provided it is
given an extra energy by some other mechanism (e.g. stellar
pulsation, interaction in a binary system, photospheric activity).
This material is likely to have, initially, a very complicated
density and velocity distribution. As the gas diffuses outwards as
a result of viscosity, the density and velocity distributions will
tend to relax and become in hydrostatic equilibrium.

Thus, it is reasonable to assume that there is a \emph{transition
region} between the photosphere (which is characterized by large
densities of the order of $10^{-10}\;\rm g\; cm^3$) and the disk
itself, which has densities at least 10 times lower. Many unknowns
exist concerning the size and physical properties
(e.g. density, temperature, velocity field) of this smooth
transition region. 

With the available data we were able to narrow down substantially the range of possible values for Achernar's model parameters, but we cannot distinguish yet between model 2 and 3, i.e., between a model in hydrostatic equilibrium and a model with enhanced scaleheights that might result from the perturbation of the gas by some photospheric mechanism. 

As shown in Figure~\ref{fig:sed}, high-precision spectropolarimetry might add invaluable information, since models 2 and 3 have different values for the size of Balmer and Pashen jumps in polarization.
Also, with simultaneous spectro-interferometry and spectropolarimetry one can 
spatially resolve the velocity field, thus establishing whether the material is in equilibrium or not. The observed residual emission profile (Figure~\ref{fig:sed}) is indicative of velocities fields more complex than the simple Keplerian rotation that was assumed in this work.

In any case, one important consequence of this work is that we have shown that with current observing techniques and state-of-the-art modeling it is already possible to study the properties of the very inner layers of the disk in Achernar and possibly other nearby Be stars. Future simultaneous spectropolarimetry, spectro-interferometry and photometry will allow us to study the properties of this region and determine, for instance, the
size of the transition region and the point at which the disk becomes in vertical hydrostatic equilibrium.

\acknowledgements{
This work was supported by FAPESP grant 04/07707-3 to ACC, FAPESP grant 01/12598-1 to AMM and NSF  grant AST-0307686 to the University of Toledo (JEB). AMM also acknowledges partial support by CNPq.
}

\clearpage

\begin{deluxetable}{lll}
\tablecaption{Fixed Stellar Parameters. \label{tab:star}}
\tablewidth{0pt}
\tablehead{
\colhead{Parameter} &
\colhead{Value} &
\colhead{Reference}}
\startdata
$R_\mathrm{pole}$ & $7.3\; R_\sun$ & This work \\
Luminosity & $3150\; L_\sun$ &  This work \\
$V_\mathrm{crit}$ & $350\;\rm km\;s^{-1}$ & \citet{vin06}  \\
inclination & $65\deg$ & \citet{car07}  \\
distance & $44.1\;\rm pc$ & \citet{per97}\\
\enddata
\end{deluxetable}


\begin{deluxetable}{cccccccccc}
\tablecaption{Model Parameters. \label{tab:models}}
\tablewidth{0pt}
\tablehead{
\colhead{Model} &
\colhead{$\Omega/\Omega_\mathrm{crit}$} &
\colhead{$\rho_0\;(\rm g\;cm{-3})$} &
\colhead{$R_\mathrm{d}$} &
\colhead{$H\;(R_\sun)$} &
\colhead{$\frac{R_\mathrm{eq}}{R_\mathrm{p}}$} &
\colhead{$\frac{T_\mathrm{p}}{T_\mathrm{eq}}$} &
\colhead{H$\alpha$ EW (\AA)} &
\colhead{$P_B$ (\%)} &
\colhead{$E_K$}
}
\startdata
1  &  0.999999  &   $1.0 \times 10^{-11}$ & 13.7 & 0.45 & 1.5 & 5.8 &
0.10 &  0.02 &  0.17 \\ 
2  &  0.999999  &   $3.8 \times 10^{-12}$ & 19.1 & 0.45 & 1.5 & 5.8 &
-0.34 & 0.10  &  0.14   \\ 
3  &  0.999999  &   $2.6 \times 10^{-12}$ & 16.1 & 1.1  & 1.5 & 5.8 &
-0.27 &  0.12  & 0.12  \\ 
4  &  0.992  &   $1.0 \times 10^{-11}$ & 13.7  &  0.45 &  1.4 &1.8 &
0.20 &  0.31 &  0.15  \\ 
5  &  0.992  &     $1.7 \times 10^{-12}$ & 13.7 & 0.91 & 1.4  &1.8 &
-0.05 &  0.15 & 0.03   \\ 
\enddata
\end{deluxetable}

\clearpage 

\begin{thebibliography}{}

\bibitem[Carciofi \& Bjorkman(2006)]{car06a} Carciofi, A.~C.,
\& Bjorkman, J.~E.\ 2006, \apj, 639, 1081

\bibitem[Carciofi et al.(2006)]{car06b} Carciofi, A.~C., Miroshnitchenko, A. S., Kusakin, A. V.,  et
al.\ 2006, \apj, 652, 1617

\bibitem[Carciofi et al.(2007)]{car07} Carciofi, A.~C.,
Magalh{\~a}es, A.~M., Leister, N.~V., Bjorkman, J.~E., \& Levenhagen,
R.~S.\ 2007, \apjl, in press

\bibitem[Cutri et al.(2003)]{2mass} Cutri, R.~M., et al.\ 
2003, The IRSA 2MASS All-Sky Point Source Catalog, NASA/IPAC Infrared 
Science Archive.~http://irsa.ipac.caltech.edu/applications/Gator/

\bibitem[Domiciano de Souza et al.(2003)]{dom03} Domiciano de Souza, A., Kervella,
P., Jankov, S., Abe, L., Vakili, F., di Folco, E., \& Paresce, F.\ 2003, \aap, 407, L47 (D03)

\bibitem[Fr{\'e}mat et al.(2005)]{fre05} Fr{\'e}mat, Y.,
Zorec, J., Hubert, A.-M., \& Floquet, M.\ 2005, \aap, 440, 305

\bibitem[Jackson et al.(2004)]{jac04} Jackson, S., MacGregor,
K.~B., \& Skumanich, A.\ 2004, \apj, 606, 1196

\bibitem[Kervella et al.(2003)]{ker03} Kervella, P., et al.\ 
2003, \procspie, 4838, 858 

\bibitem[Kervella \& Domiciano de Souza(2006)]{ker06}
Kervella, P., \& Domiciano de Souza, A.\ 2006, \aap, 453, 1059

\bibitem[Kervella \& Domiciano de Souza(2007)]{ker07}
Kervella, P., \& Domiciano de Souza, A.\ 2007, \aap, 474, L49

\bibitem[Kurucz (1994)]{kur94} Kurucz, R. L. 1994, Kurucz CD-ROM No. 19, 20, 21, Cambridge Mass.: Smithsonian Astrophysical Observatory

\bibitem[Meilland et al.(2007)]{mei07} Meilland, A., Stee, P., Vannier, M. et al. \ 2007, \aap, 464, 59

\bibitem[Perryman et al.(1997)]{per97} Perryman, M. A. C., Lindegren, L., Kovalevsky, J., et al. \ 1997,
\aap, 323, L49

\bibitem[Vinicius et al.(2006)]{vin06} Vinicius, M.~M.~F.,
Zorec, J., Leister, N.~V., \& Levenhagen, R.~S.\ 2006, \aap, 446, 643

\bibitem[Zacharias et al.(2005)]{nomad} Zacharias, N., Monet, 
D.~G., Levine, S.~E., Urban, S.~E., Gaume, R., \& Wycoff, G.~L.\ 2005, 
VizieR Online Data Catalog, 1297, 0 

\bibitem[von Zeipel(1924)]{von24} von Zeipel, H.\ 1924,
\mnras, 84, 665

\end{thebibliography}
\end{document}